\documentclass[prl,twocolumn,showpacs,superscriptaddress,floatfix,longbibliography]{revtex4-1}
\usepackage{graphicx,amsfonts,amssymb,amsmath,xspace,dsfont}
\usepackage[colorlinks=true,citecolor=green,linkcolor=red,urlcolor=blue]{hyperref}
\usepackage{subfigure}
\usepackage{calligra}

\usepackage{color}
\definecolor{g}{rgb}{.1,0.4,.1} 
\definecolor{b}{rgb}{0,0.2,1}
\definecolor{rouge}{rgb}{0.82,0.,0.}
\definecolor{vert}{rgb}{0.,0.82,0.}
\definecolor{orange}{rgb}{1,0.5,0.}
\definecolor{bleu}{rgb}{0.,0.,0.82}
\definecolor{m}{rgb}{0.82,0.,0.82}
\definecolor{vert2}{rgb}{0.,0.5,0.}
\definecolor{rougeclair}{rgb}{1.0,0.7,0.7}

\newcommand{\beq}{\begin{equation}}
\newcommand{\be}{\begin{equation}}
\newcommand{\beqn}{\begin{eqnarray}}
\newcommand{\eeq}{\end{equation}}
\newcommand{\ee}{\end{equation}}
\newcommand{\eeqn}{\end{eqnarray}}

\newcommand{\q}{{\bf q}}

\newcommand{\bem}{\begin{pmatrix}}

\newcommand{\eem}{\end{pmatrix}}

\newlength{\ldag}
\newcommand{\Pp}{B_p}

\begin{document}

\title{Partition function of the Levin-Wen model}

\author{Julien Vidal}
\email{vidal@lptmc.jussieu.fr}
\affiliation{Sorbonne Universit\'e, CNRS, Laboratoire de Physique Th\'eorique de la Mati\`ere Condens\'ee, LPTMC, F-75005 Paris, France}

\begin{abstract}

Using a description of the Levin-Wen model excitations in terms of Wilson lines, we compute the degeneracy of the energy levels for any input anyon theory and for any trivalent graph embedded on any (orientable) compact surface. This result allows one to obtain the finite-size and finite-temperature partition function and to show that there are no thermal phase transitions.
\end{abstract}

\maketitle

%
%
%
Topological quantum phases of matter have been intensively studied following the discovery of the fractional quantum Hall effect \cite{Tsui82}. In two dimensions, these phases are characterized by exotic emergent excitations known as anyons \cite{Leinaas77,Wilczek82} whose nontrivial braiding statistics have been recently probed experimentally \cite{Bartolomei20,Nakamura20}. Anyons are deeply related to the concept of topological order developed in the context of high-$T_c$ superconductivity \cite{Wen90_1,Wen17}. An important property of topologically ordered systems is the dependence of the  ground-state degeneracy with respect to the surface topology. The robustness of this degeneracy against local perturbations makes anyons promising candidates for topological quantum computation \cite{Kitaev03,Freedman03,Nayak08,Wang_book,Preskill_HP}. 
At low energies, topologically ordered systems are described by topological quantum field theories establishing a link between the ground-state degeneracy and the knot invariants~ \cite{Witten89}. 

Topological phases can be split into two families: Chiral phases which break time-reversal symmetry and sustain gapless edge modes and achiral (doubled) phases which are time-reversal symmetric and gapped. If chiral phases are definitely of great relevance, for instance, to understand fractional quantum Hall effect, achiral phases are also of interest since they can be realized in several microscopic models such as the toric code~\cite{Kitaev03} or the Levin-Wen model \cite{Levin05} which is the focus of the present study. This model, also known as the string-net model, can generate any doubled achiral topological phase. Furthermore, relaxing some constraints imposed in its original formulation, the generalized string-net model discussed in Refs.~\cite{Hahn20,Lin21,Wolf_thesis} allows to generate some topological phases that are not time-reversal symmetric.  In these past years, the Levin-Wen model has been of considerable interest, notably to study the quantum phase transitions driven by a string tension \cite{Gils09_1,Gils09_3,Ardonne11,Burnell11_2,Schulz13,Schulz14,Dusuel15,Schulz15,Schulz16_2,Mariens17,Schotte19,Ritz21}. However, several properties of the original (unperturbed) model are already fascinating. For instance, as discussed in Refs.~\cite{Kadar10,Burnell10,Burnell11_1,Kirillov11}, the zero-temperature partition function (ground-state degeneracy) is known to be related to the Turaev-Viro \cite{Turaev92} and Witten-Reshitikhin-Turaev invariants \cite{Witten89,Reshetikhin91}.

Widely inspired by lattice gauge theories, the Levin-Wen model has two kinds of excitations: Charge excitations associated with  violations of vertex constraints and flux excitations associated with violations of plaquette constraints. As suggested early in Ref.~\cite{Simon13}, excitations of flux type (dubbed fluxons) can be thought of as Wilson lines through the plaquettes of the lattice. This description has been very useful to compute the degeneracy of the energy levels for some specific anyon \mbox{theories~\cite{Schulz13,Schulz14,Schulz15,Hu18}}. 

In this Letter, we use this fluxon picture to exactly compute the degeneracy of the energy levels in the Levin-Wen model for {\it any anyon (modular) theory} and for {\it any trivalent graph embedded on any orientable compact surface}. This degeneracy is found to depend only on the quantum dimensions of the anyons and on the genus of the surface considered. We also give the results for the cylindrical geometry which is relevant for the two-leg ladder~\cite{Gils09_1,Gils09_3,Ardonne11,Morampudi14,Schulz15,Vidal18}. Next, we compute exactly the {\it finite-temperature} partition function for any {\it finite-size} system. The zero- and the infinite-temperature limits directly give simple expressions of the ground-state degeneracy and of the Hilbert space dimension, respectively. In the thermodynamical limit, the partition function still depends on the surface topology but this is not the case for thermodynamical quantities such as heat capacity. Finally, we show that there are no finite-temperature phase transitions in this two-dimensional microscopic lattice model. \\ 

%
%
\noindent\emph{The Levin-Wen model in a nutshell} ---
%
The Levin-Wen model \cite{Levin05} can be defined on any two-dimensional trivalent graph. Microscopic degrees of freedom are strings associated with the links of this graph. The Hilbert space~${\mathcal H}$ is spanned by all string configurations satisfying some branching rules at each vertex. These branching rules directly stem from the fusion rules of the considered input theory. In other words, a vertex configuration $(a,b,c)$ is allowed iff $c$ belongs to the fusion product $a\times b$ (at this stage, we omit the possible orientation of the string that becomes crucial for non self-dual theories). Violations of these rules correspond to charge excitations that are not considered here. 
In the following, we consider unitary modular tensor categories (UMTCs) as input theory.  As explained, for instance, in Refs.~\cite{Rowell09,Bonderson_thesis,Wang_book}, UMTCs are the proper mathematical objects to describe anyons. A~UMTC of rank $n$ is essentially defined by three quantities \cite{Rowell09}~: A set $\mathcal N$ of fusion matrices $N_{i=1\dots n}$ which encode the fusion rules between the $n$ different strings, the modular $S$ matrix, and the $T$ matrix~\cite{Rowell09,Bonderson_thesis,Wang_book}.
\newpage

\begin{figure}[t]
\includegraphics[width=0.8\columnwidth]{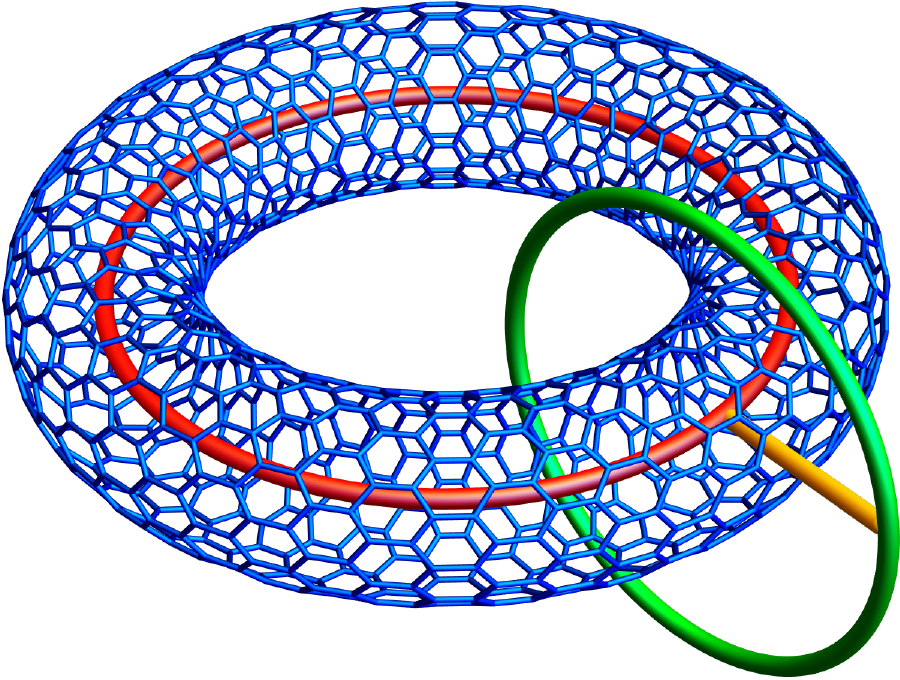}
\caption{Illustration of a single-fluxon excitation (yellow) associated with a red (green) noncontractible loop inside (outside) the torus $(g=1)$. For an input UMTC with $n$ different strings, these loops can each take $n$ different values leading to a ground-state degeneracy \mbox{$\mathcal{D}_0=n^2$} for the torus [see Eq.~(\ref{eq:gse_deg})]. Here we choose the honeycomb lattice as an example of trivalent graph.}
\label{fig:fluxon}
\end{figure}

The Hamiltonian of the Levin-Wen model is given \mbox{by \cite{Levin05}} 
%
%
\begin{equation}
H= - \sum_{p=1}^{N_{\rm p}} \Pp, 
\label{eq:ham}
\end{equation}
%
%
where $\Pp$'s are local commuting projectors acting on the plaquette $p$ ($N_{\rm p}$ is the total number of plaquettes in the system). As explained in Ref.~\cite{Levin05}, matrix elements of $\Pp$ in the original link basis are expressed in terms of $F$ symbols given by the input category. However, for our purpose, we will use a different approach and, following Ref.~\cite{Simon13}, interpret $\Pp$ as an operator that projects onto the trivial flux (vacuum) in the plaquette $p$. In this representation in terms of flux lines (also known as fluxons or Wilson lines), the possibly degenerate ground states are all states with no flux in the plaquettes and excitations are simply lines piercing the lattice (see Fig.~\ref{fig:fluxon}). The resulting doubled achiral topological phase ${\rm D} (\mathcal C)=( \mathcal C, \overline{\mathcal C})$ consists of two copies of the input UMTC~$\mathcal{C}$ with opposite chiralities \cite{Levin05}, and excitations can be labeled by $(s,s')$, where $s$ and $s'$ are elements of  $\mathcal{C}$ and $\overline{\mathcal C}$, respectively. However, in the absence of charge excitations, elementary excitations correspond to $s=s'$. As a direct consequence, the Wilson lines associated with $(s,\overline{s})$ obey the fusion rules of the input theory $\mathcal{C}$. At the end of the day, counting degeneracies of energy levels simply amounts to dealing with the fusion rules. \\ 
 
%
\noindent\emph{Spectrum degeneracy} ---
%
To compute the energy level degeneracies, we need to introduce a few ingredients. Let us consider a general input UMTC $\mathcal{C}$ with $n$ different strings. For simplicity, we assume here that $\mathcal{C}$ is multiplicity free, i.e., that the fusion coefficients are either 0 or 1. Following Rowell {\it et al.} \cite{Rowell09}, we define the fusion matrices $N_i$ through the symmetric unitary $n\times n $ matrix $S$ by using the Pasquier-Verlinde formula \cite{Pasquier87,Verlinde88} which can be written as
%
%
\begin{equation}
N_i= S \Lambda_i S^\dagger, 
\label{eq:Verlinde}
\end{equation}
%
%
where $\Lambda_i$ is a diagonal matrix whose entries are given by $(\Lambda_{i})_{j,k}=\delta_{j,k} \, S_{i,j}/S_{1,j}$.  By convention, the trivial string (vacuum) is chosen as the string $j=1$. Let us remind that $S^2=C$ where $C$ is the charge conjugation matrix [\mbox{$C_{j,k}=1$} if $j$ and $k$ are conjugate (i.e., they can fuse to 1) and 0 otherwise].
For each string $j$, the quantum dimension is defined as \mbox{$d_j=S_{1,j}/S_{1,1}$}. 
We emphasize that $N_i$'s and $d_i$'s obey the associative commutative fusion algebra of $\mathcal{C}$ (see Ref.~\cite{Gannon05} for more details). As such, fusion matrices mutually commute and are simultaneously diagonalizable as anticipated from Eq.~(\ref{eq:Verlinde}). Thus, for our purpose, only the $S$ matrix matters.

By construction,  $(N_i)_{j,k}=1$ if $k \in i\times j$, and 0 otherwise. Hence, if one fuses together a set of $l_i$ strings of type $i$ (with $i=1,\dots,n$), the number of ways to obtain the string $k$ is given by: 
%
%
\begin{eqnarray}
\mathcal{F}^{\{l_i\}}_k&=& \Bigg( \prod_{i=1}^{n} N_i^{l_i} \Bigg) _{1,k}, \\
&=& \Bigg[ S \Bigg(\prod_{i=1}^{n} \Lambda_i^{l_i} \Bigg) S^\dagger \Bigg]_{1,k},\\
&=& \sum_{j=1}^{n} S_{1,j} \prod_{i=1}^{n} \left( \frac{S_{i,j}}{S_{1,j}} \right)^{l_i} S^\dagger_{j,k}, 
\label{eq:fusion}
\end{eqnarray}
%
%
where we used  the Verlinde formula (\ref{eq:Verlinde}) to obtain the last result. 
%
%
%
%

\begin{figure}[t]
\includegraphics[width=0.8\columnwidth]{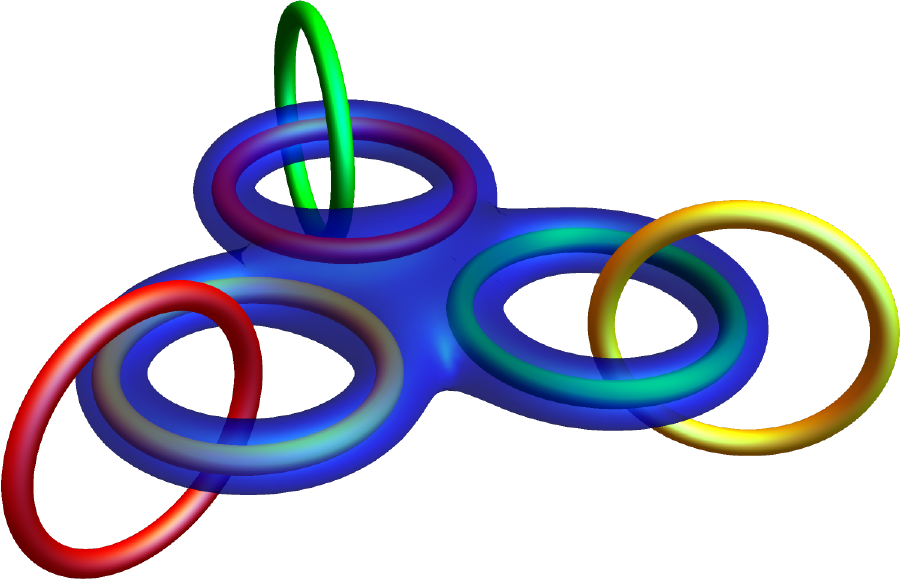}
\caption{A possible choice of independent inner and outer noncontractible loops for a 3-torus ($g=3$). For non-Abelian theories, these loops can fuse in several ways into the vacuum string $1$ and $\mathcal{D}_0$ is given by Eq.~(\ref{eq:gse_deg}). In the Abelian case ($S_{1,j}=1/\sqrt{n}, \forall j$), the ground-state degeneracy on a genus $g$ surface reduces to $\mathcal{D}_0=n^{2g}$. For clarity, the trivalent graph is not displayed on the surface. }
\label{fig:3_torus}
\end{figure}

Next, one must specify the boundary conditions. In a first step, let us consider a  trivalent graph embedded on a compact orientable surface of genus $g$.  Then, one has $g$ independent noncontractible loops inside as well as outside the surface (see Fig.~\ref{fig:3_torus} for $g=3$).   
To determine the degeneracy $\mathcal D_q$ of the $q$-th energy level on such a surface, one has to compute the number of possibilities to have any $q$ nontrivial ($j\neq 1$) fluxons piercing the lattice and the number of ways to connect the overall fluxon inside and outside the surface to all possible inner and outer overall fluxes while respecting the fusion rules. 

Using Eq.~(\ref{eq:fusion}), one can compute the total number of ways to fuse $g$ noncontractible loops into a given \mbox{string $k$},
%
%
\begin{eqnarray}
\mathcal{G}_{g,k}&=& 
\sum_{\{ l_i\}/g} \sum_{k_1=1}^n \sum_{k_2=1}^n \sum_{j=1}^{n}
\mathcal{F}^{\{l_i\}}_{k_1} \mathcal{F}^{\{l_i\}}_{k_2} S_{1,j} \frac{S_{k1,j}}{S_{1,j}} \frac{S_{k_2,j}}{S_{1,j}} S^\dagger_{j,k} \nonumber \\
&&\left(\hspace{-1mm}\begin{array}{c} g \\l_1\end{array}\hspace{-1mm}\right) 
\left(\hspace{-1mm}\begin{array}{c} g-l_1 \\l_2\end{array}\hspace{-1mm}\right)\dots
\left(\hspace{-1mm}\begin{array}{c} g-l_1-l_2-... \, l_{n-1} \\ l_n\end{array}\hspace{-1mm}\right), \label{eq:G} \\
&=&\sum_{j=1}^n S_{1,j}^{1-2g} S^\dagger_{j,k},
\end{eqnarray} 
%
%
where the first sum $\sum_{\{ l_i\}/g}$ is performed over all possible choices of $\{ l_i\}$ such that $\sum_{i=1}^n l_i=g$ and where the product of binomials accounts for all possible permutations of the $g$ strings considered. Equation (\ref{eq:G}) can be interpreted as the result of $g$ strings (one in each hole) fusing in $k_1$ and $k_2$ ``above" and ``below" the $g$ holes of the surface and imposing that these two strings $k_1$ and $k_2$ fuse \mbox{into $k$}.
%
%
%
%
%

As a first result, one then gets a simple expression of the ground-state degeneracy
%
%
\begin{equation}
{\mathcal D}_0=(\mathcal{G}_{g,1})^2= \Bigg(\sum_{j=1}^n S_{1,j}^{\chi}\Bigg)^2,
\label{eq:gse_deg}
\end{equation}
%
%
where we introduced the Euler-Poincar\'e characteristic  \mbox{$\chi=N_{\rm v}-N_{\rm l}+N_{\rm p}$}, where $N_{\rm v}, N_{\rm l}$, and $N_{\rm p}$ denote the total number of  vertices, links, and plaquettes, respectively. For a compact surface of genus $g$, one has $\chi=2-2g$. This simple form is reminiscent of the doubled structure of the topological phase emerging from the Levin-Wen model. Indeed, as expected, $\mathcal D_0$ can be identified as the square of the quantity derived in Ref.~\cite{Barkeshli09} for non-doubled theories (see also Ref.\cite{Verlinde88} for an older derivation in the context of conformal field theories). 


As explained in Refs~\cite{Kadar10,Burnell10,Burnell11_1,Kirillov11}, $\mathcal D_0$ is exactly the Turaev-Viro invariant \cite{Turaev92}. 
Hence, Eq.~(\ref{eq:gse_deg}) can be considered as an explicit form of this invariant for any UMTC. For a surface with $g=0$, one gets $\mathcal D_0=1$ (unique ground state), as anticipated in Ref.~\cite{Hu12}. For a surface with \mbox{$g=1$}, we find that $\mathcal D_0=n^2$ so that, in this case, the ground-state degeneracy only depends on the number of strings. More generally, $\mathcal D_0$ on a torus is given by the number of particles contained in the Drinfel'd double ${\rm D} (\mathcal C)$ \cite{Drinfeld87}. By contrast, for $g \geqslant 2$, $\mathcal D_0$ depends on the quantum dimensions $d_j$'s which can be different for a fixed~$n$. For non-modular theories, a general formula of~$\mathcal D_0$ in terms of $6j$ symbols can be found in Ref.~\cite{Hu12}.

Then, according to the above discussion, one gets: 
%
%
\begin{eqnarray}
\mathcal{D}_q&=& 
\left(\hspace{-1mm}
\begin{array}{c}
N_{\mathrm p}
\\
q
\end{array}
\hspace{-1mm}
\right) \sum_{\{ l_i\}/q}
\sum_{k_1=1}^n \sum_{k_2=1}^n 
\mathcal{F}^{\{l_i\}}_{k_1} \mathcal{F}^{\{l_i\}}_{k_2} \mathcal{G}_{g,k_1} \mathcal{G}_{g,k_2} \nonumber\\
&&
\left(\hspace{-1mm}\begin{array}{c}q \\l_2\end{array}\hspace{-1mm}\right) 
\left(\hspace{-1mm}\begin{array}{c}q-l_2 \\l_3\end{array}\hspace{-1mm}\right)\dots
 \left(\hspace{-1mm}\begin{array}{c}q-l_2-l_3-... \, l_{n-1} \\ l_n\end{array}\hspace{-1mm}\right),
 \label{eq:deg_f}
\end{eqnarray} 
%
%
where the first binomial takes into account all the possibilities to choose the position of the $q$ nontrivial fluxons among the $N_{\rm p}$ plaquettes. The last product of binomials accounts for all possible permutations of the different strings in the set of these $q$ nontrivial fluxons. The first sum is performed over all possible choices of $\{ l_i\}$ such that $\sum_{i=2}^n l_i=q$ (here the case $i=1$ is discarded since it does not correspond to an excitation).  The last two sums are performed over the possible values of $k_1$ and $k_2$ which are the inner and outer fluxons, respectively. 

After a few algebra (see Ref.~\cite{Supp_Mat} for details), the degeneracy of the $q$-fluxon states boils down to
%
%
\begin{equation}
\mathcal{D}_q=
\left(\hspace{-1mm}
\begin{array}{c}
N_{\mathrm p}
\\
q
\end{array}
\hspace{-1mm}
\right) (-1)^q  \Bigg\{ {\mathcal D}_0+\sum_{j=1}^n S_{1,j}^{2 \chi}  \bigg[\Big(1-S_{1,j}^{-2}\Big)^q-1 \bigg]\Bigg\},  \quad 
 \label{eq:deg_g}
\end{equation} 
%
%
which is the main result of this Letter. One can easily check that this general form reproduces all known results spread in the literature for Fibonacci \cite{Schulz13,Hu14} and Ising theories \cite{Schulz14} on the torus ($g=1$). 

Interestingly, Eq.~(\ref{eq:deg_g}) is also valid for non-compact surfaces such as the cylinder ($\mathcal{D}_0=n$, $\chi=0$) or the plane ($\mathcal{D}_0=1$, $\chi=1$). Note that the cylindrical geometry is relevant for the two-leg ladder with periodic boundary conditions studied for some specific theories in \mbox{Refs.~\cite{Gils09_1,Gils09_3,Ardonne11,Morampudi14,Schulz15,Vidal18}}. Degeneracies of the energy levels for the Fibonacci theory on various topologies are given in Ref.~\cite{Supp_Mat} for illustration.\\

%
\noindent\emph{Partition function} ---
%
Using Eq.~(\ref{eq:deg_g}), it is straightforward to compute the partition function of the model. Keeping in mind that the energy of any $q$-fluxon state is \mbox{$E_q=-N_{\rm p}+q$}, one gets
%
%
\begin{eqnarray}
Z&=& \sum_{q=0}^{N_{\rm p}} \mathcal{D}_q \:  {\rm e}^{-\beta E_q},\\
&=& \big({\rm e}^{\beta}-1\big)^{N_{\rm p}} \nonumber \\
&&
 \Bigg\{ 
{\mathcal D}_0 +\sum_{j=1}^n S_{1,j}^{2 \chi}
\Bigg[
\bigg(
1+\frac{S_{1,j}^{-2}}{{\rm e}^{\beta}-1}  
\bigg)^{N_{\rm p}}-1
\Bigg]
\Bigg\},\qquad 
\label{eq:Z}
\end{eqnarray} 
%
%
where  $\beta=1/T$ is the inverse temperature. From the partition function, one can easily compute the dimension of the Hilbert space by taking the infinite-temperature  limit. One then obtains:
%
%
\begin{equation}
\dim {\mathcal H} = \lim_{\beta \rightarrow 0} Z=
\sum_{j=1}^n
S_{1,j}^{-N_{\rm v}}.
\label{eq:dimH}
\end{equation} 
%
%
Here we used the fact that for any trivalent graph, one has  $N_{\rm l}=3N_{\rm v}/2$  \{see also  Refs.~\cite{Gils09_1_s,Gils09_1,Gils09_3,Simon13,Hermanns14} for some alternative derivations of Eq.~(\ref{eq:dimH})\}. 
In the thermodynamical limit, keeping in mind that $S_{1,j}\geqslant S_{1,1}$, one recovers the following well-known result:
%
%
\begin{equation}
\dim {\mathcal H} \underset{N_{\rm v} \rightarrow \infty}{\sim} D^{N_{\rm v}},
\end{equation} 
%
%
which gives a physical interpretation to the  total quantum dimension $D=1/S_{1,1}$.\\

%
\noindent\emph{Specific heat} ---
%
The partition function $Z$ provides  direct access to most thermodynamical quantities. To analyze the possibility of a temperature-driven phase transition in the Levin-Wen model, we consider the specific heat (per plaquette)
%
%
\begin{equation}
c =\frac{\beta^2}{N_{\mathrm p}}\frac{\partial^2 \ln Z}{\partial \beta^2},
\end{equation}
%
%
which must be singular at the transition temperature, if any. Using the partition function given in Eq.~(\ref{eq:Z}) and taking the thermodynamical limit  ($N_{\rm p} \rightarrow \infty$), one finds:
%
%
\begin{equation}
c=\frac{\mathrm{e}^{\beta} \,\beta^2 (D^2-1)}{(D^2-1+\mathrm{e}^{\beta})^2}.
\end{equation}
%
%
As could be expected, $c$ is not sensitive to the topology of the system, i.e., it is independent of $\chi$. More interestingly, we find that $c$ only depends on the total quantum dimension $D$. Hence, two completely different theories [such as, e.g.,  Ising (non-Abelian and self-dual) and $\mathbb{Z}_4$ (Abelian and non-self-dual) for which $D=2$], may have the same $c$. But, the most important result is that, for any $D$, $c$ is a smooth function of $\beta$ indicating the absence of thermal phase transitions in the Levin-Wen model. \\

%
\noindent\emph{Outlook} ---
%
Using solely fusion algebra ($S$matrix), we determined the spectrum degeneracy of the Levin-Wen model for any orientable compact surface and for any input anyon theory. This allowed us to compute the partition function for any finite-size system at any finite temperature and to prove that there are no thermal phase transition in this model. 

A natural extension of this work would consist in analyzing non-modular theories for which $\det S=0$. In this case, the Wilson-line picture used here fails and one must rather use the formalism developed in the context of quantum double models \cite{Kitaev03}. We conjecture that Eq.~(\ref{eq:deg_g}) is still valid provided (i) one replaces the noninvertible $S$ matrix by the unitary matrix $\widetilde{S}$ that simultaneously diagonalizes all fusion matrices and (ii) one sets the proper value of $\mathcal{D}_0$ which is not given by Eq.~(\ref{eq:gse_deg}) anymore. Another perspective is to extend this calculation to theories with multiplicities.
Finally, we emphasize that the knowledge of the partition function should deepen our understanding of topologically ordered systems at finite temperatures as recently discussed in Ref.~\cite{Weinstein19}.  

 
\acknowledgments 
I am very grateful to S. Dusuel for many valuable insights and discussions at the early stages of this work.


\begin{thebibliography}{52}%
\makeatletter
\providecommand \@ifxundefined [1]{%
 \@ifx{#1\undefined}
}%
\providecommand \@ifnum [1]{%
 \ifnum #1\expandafter \@firstoftwo
 \else \expandafter \@secondoftwo
 \fi
}%
\providecommand \@ifx [1]{%
 \ifx #1\expandafter \@firstoftwo
 \else \expandafter \@secondoftwo
 \fi
}%
\providecommand \natexlab [1]{#1}%
\providecommand \enquote  [1]{``#1''}%
\providecommand \bibnamefont  [1]{#1}%
\providecommand \bibfnamefont [1]{#1}%
\providecommand \citenamefont [1]{#1}%
\providecommand \href@noop [0]{\@secondoftwo}%
\providecommand \href [0]{\begingroup \@sanitize@url \@href}%
\providecommand \@href[1]{\@@startlink{#1}\@@href}%
\providecommand \@@href[1]{\endgroup#1\@@endlink}%
\providecommand \@sanitize@url [0]{\catcode `\\12\catcode `\$12\catcode
  `\&12\catcode `\#12\catcode `\^12\catcode `\_12\catcode `\%12\relax}%
\providecommand \@@startlink[1]{}%
\providecommand \@@endlink[0]{}%
\providecommand \url  [0]{\begingroup\@sanitize@url \@url }%
\providecommand \@url [1]{\endgroup\@href {#1}{\urlprefix }}%
\providecommand \urlprefix  [0]{URL }%
\providecommand \Eprint [0]{\href }%
\providecommand \doibase [0]{http://dx.doi.org/}%
\providecommand \selectlanguage [0]{\@gobble}%
\providecommand \bibinfo  [0]{\@secondoftwo}%
\providecommand \bibfield  [0]{\@secondoftwo}%
\providecommand \translation [1]{[#1]}%
\providecommand \BibitemOpen [0]{}%
\providecommand \bibitemStop [0]{}%
\providecommand \bibitemNoStop [0]{.\EOS\space}%
\providecommand \EOS [0]{\spacefactor3000\relax}%
\providecommand \BibitemShut  [1]{\csname bibitem#1\endcsname}%
\let\auto@bib@innerbib\@empty
\bibitem [{\citenamefont {Tsui}\ \emph {et~al.}(1982)\citenamefont {Tsui},
  \citenamefont {Stormer},\ and\ \citenamefont {Gossard}}]{Tsui82}%
  \BibitemOpen
  \bibfield  {author} {\bibinfo {author} {\bibfnamefont {D.~C.}\ \bibnamefont
  {Tsui}}, \bibinfo {author} {\bibfnamefont {H.~L.}\ \bibnamefont {Stormer}}, \
  and\ \bibinfo {author} {\bibfnamefont {A.~C.}\ \bibnamefont {Gossard}},\
  }\bibfield  {title} {\enquote {\bibinfo {title} {{Two-Dimensional
  Magnetotransport in the Extreme Quantum Limit}},}\ }\href {\doibase
  10.1103/PhysRevLett.48.1559} {\bibfield  {journal} {\bibinfo  {journal}
  {Phys. Rev. Lett.}\ }\textbf {\bibinfo {volume} {48}},\ \bibinfo {pages}
  {1559} (\bibinfo {year} {1982})}\BibitemShut {NoStop}%
\bibitem [{\citenamefont {Leinaas}\ and\ \citenamefont
  {Myrheim}(1977)}]{Leinaas77}%
  \BibitemOpen
  \bibfield  {author} {\bibinfo {author} {\bibfnamefont {J.~M.}\ \bibnamefont
  {Leinaas}}\ and\ \bibinfo {author} {\bibfnamefont {J.}~\bibnamefont
  {Myrheim}},\ }\bibfield  {title} {\enquote {\bibinfo {title} {{On the theory
  of identical particles}},}\ }\href {\doibase 10.1007/BF02727953} {\bibfield
  {journal} {\bibinfo  {journal} {Nuovo Cim. B}\ }\textbf {\bibinfo {volume}
  {37}},\ \bibinfo {pages} {1} (\bibinfo {year} {1977})}\BibitemShut {NoStop}%
\bibitem [{\citenamefont {Wilczek}(1982)}]{Wilczek82}%
  \BibitemOpen
  \bibfield  {author} {\bibinfo {author} {\bibfnamefont {F.}~\bibnamefont
  {Wilczek}},\ }\bibfield  {title} {\enquote {\bibinfo {title} {{Quantum
  Mechanics of Fractional-Spin Particles}},}\ }\href {\doibase
  10.1103/PhysRevLett.49.957} {\bibfield  {journal} {\bibinfo  {journal} {Phys.
  Rev. Lett.}\ }\textbf {\bibinfo {volume} {49}},\ \bibinfo {pages} {957}
  (\bibinfo {year} {1982})}\BibitemShut {NoStop}%
\bibitem [{\citenamefont {Bartolomei}\ \emph {et~al.}(2020)\citenamefont
  {Bartolomei}, \citenamefont {Kumar}, \citenamefont {Bisognin}, \citenamefont
  {Marguerite}, \citenamefont {Berroir}, \citenamefont {Bocquillon},
  \citenamefont {Pla\c{c}ais}, \citenamefont {Cavanna}, \citenamefont {Dong},
  \citenamefont {Gennser}, \citenamefont {Jin},\ and\ \citenamefont
  {F\`eve}}]{Bartolomei20}%
  \BibitemOpen
  \bibfield  {author} {\bibinfo {author} {\bibfnamefont {H.}~\bibnamefont
  {Bartolomei}}, \bibinfo {author} {\bibfnamefont {M.}~\bibnamefont {Kumar}},
  \bibinfo {author} {\bibfnamefont {R.}~\bibnamefont {Bisognin}}, \bibinfo
  {author} {\bibfnamefont {A.}~\bibnamefont {Marguerite}}, \bibinfo {author}
  {\bibfnamefont {J.-M.}\ \bibnamefont {Berroir}}, \bibinfo {author}
  {\bibfnamefont {E.}~\bibnamefont {Bocquillon}}, \bibinfo {author}
  {\bibfnamefont {B.}~\bibnamefont {Pla\c{c}ais}}, \bibinfo {author}
  {\bibfnamefont {A.}~\bibnamefont {Cavanna}}, \bibinfo {author} {\bibfnamefont
  {Q.}~\bibnamefont {Dong}}, \bibinfo {author} {\bibfnamefont {U.}~\bibnamefont
  {Gennser}}, \bibinfo {author} {\bibfnamefont {Y.}~\bibnamefont {Jin}}, \ and\
  \bibinfo {author} {\bibfnamefont {G.}~\bibnamefont {F\`eve}},\ }\bibfield
  {title} {\enquote {\bibinfo {title} {{Fractional statistics in anyon
  collisions}},}\ }\href {\doibase 10.1126/science.aaz5601} {\bibfield
  {journal} {\bibinfo  {journal} {Science}\ }\textbf {\bibinfo {volume}
  {368}},\ \bibinfo {pages} {173} (\bibinfo {year} {2020})}\BibitemShut
  {NoStop}%
\bibitem [{\citenamefont {Nakamura}\ \emph {et~al.}(2020)\citenamefont
  {Nakamura}, \citenamefont {Liang}, \citenamefont {Gardner},\ and\
  \citenamefont {Manfra}}]{Nakamura20}%
  \BibitemOpen
  \bibfield  {author} {\bibinfo {author} {\bibfnamefont {J.}~\bibnamefont
  {Nakamura}}, \bibinfo {author} {\bibfnamefont {S.}~\bibnamefont {Liang}},
  \bibinfo {author} {\bibfnamefont {G.~C.}\ \bibnamefont {Gardner}}, \ and\
  \bibinfo {author} {\bibfnamefont {M.~J.}\ \bibnamefont {Manfra}},\ }\bibfield
   {title} {\enquote {\bibinfo {title} {{Direct observation of anyonic braiding
  statistics}},}\ }\href {\doibase 10.1038/s41567-020-1019-1} {\bibfield
  {journal} {\bibinfo  {journal} {Nat. Phys.}\ }\textbf {\bibinfo {volume}
  {16}},\ \bibinfo {pages} {931} (\bibinfo {year} {2020})}\BibitemShut
  {NoStop}%
\bibitem [{\citenamefont {Wen}(1990)}]{Wen90_1}%
  \BibitemOpen
  \bibfield  {author} {\bibinfo {author} {\bibfnamefont {X.-G.}\ \bibnamefont
  {Wen}},\ }\bibfield  {title} {\enquote {\bibinfo {title} {Topological orders
  in rigid states},}\ }\href {\doibase 10.1142/S0217979290000139} {\bibfield
  {journal} {\bibinfo  {journal} {Int. J. Mod. Phys. B}\ }\textbf {\bibinfo
  {volume} {04}},\ \bibinfo {pages} {239} (\bibinfo {year} {1990})}\BibitemShut
  {NoStop}%
\bibitem [{\citenamefont {Wen}(2017)}]{Wen17}%
  \BibitemOpen
  \bibfield  {author} {\bibinfo {author} {\bibfnamefont {X.-G.}\ \bibnamefont
  {Wen}},\ }\bibfield  {title} {\enquote {\bibinfo {title} {{Zoo of
  quantum-topological phases of matter}},}\ }\href {\doibase
  10.1103/RevModPhys.89.041004} {\bibfield  {journal} {\bibinfo  {journal}
  {Rev. Mod. Phys.}\ }\textbf {\bibinfo {volume} {89}},\ \bibinfo {pages}
  {041004} (\bibinfo {year} {2017})}\BibitemShut {NoStop}%
\bibitem [{\citenamefont {Kitaev}(2003)}]{Kitaev03}%
  \BibitemOpen
  \bibfield  {author} {\bibinfo {author} {\bibfnamefont {A.}~\bibnamefont
  {Kitaev}},\ }\bibfield  {title} {\enquote {\bibinfo {title} {{Fault-tolerant
  quantum computation by anyons}},}\ }\href {\doibase
  10.1016/S0003-4916(02)00018-0} {\bibfield  {journal} {\bibinfo  {journal}
  {Ann. Phys. (NY)}\ }\textbf {\bibinfo {volume} {303}},\ \bibinfo {pages} {2}
  (\bibinfo {year} {2003})}\BibitemShut {NoStop}%
\bibitem [{\citenamefont {Freedman}\ \emph {et~al.}(2003)\citenamefont
  {Freedman}, \citenamefont {Kitaev}, \citenamefont {Larsen},\ and\
  \citenamefont {Wang}}]{Freedman03}%
  \BibitemOpen
  \bibfield  {author} {\bibinfo {author} {\bibfnamefont {M.~H.}\ \bibnamefont
  {Freedman}}, \bibinfo {author} {\bibfnamefont {A.}~\bibnamefont {Kitaev}},
  \bibinfo {author} {\bibfnamefont {M.J.}\ \bibnamefont {Larsen}}, \ and\
  \bibinfo {author} {\bibnamefont {Wang}},\ }\bibfield  {title} {\enquote
  {\bibinfo {title} {{Topological quantum computation}},}\ }\href {\doibase
  10.1090/S0273-0979-02-00964-3} {\bibfield  {journal} {\bibinfo  {journal}
  {Bull. Amer. Math. Soc.}\ }\textbf {\bibinfo {volume} {40}},\ \bibinfo
  {pages} {31} (\bibinfo {year} {2003})}\BibitemShut {NoStop}%
\bibitem [{\citenamefont {Nayak}\ \emph {et~al.}(2008)\citenamefont {Nayak},
  \citenamefont {Simon}, \citenamefont {Stern}, \citenamefont {Freedman},\ and\
  \citenamefont {Sarma}}]{Nayak08}%
  \BibitemOpen
  \bibfield  {author} {\bibinfo {author} {\bibfnamefont {C.}~\bibnamefont
  {Nayak}}, \bibinfo {author} {\bibfnamefont {S.~H.}\ \bibnamefont {Simon}},
  \bibinfo {author} {\bibfnamefont {A.}~\bibnamefont {Stern}}, \bibinfo
  {author} {\bibfnamefont {M.}~\bibnamefont {Freedman}}, \ and\ \bibinfo
  {author} {\bibfnamefont {S.~Das}\ \bibnamefont {Sarma}},\ }\bibfield  {title}
  {\enquote {\bibinfo {title} {{Non-Abelian anyons and topological quantum
  computation}},}\ }\href {\doibase 10.1103/RevModPhys.80.1083} {\bibfield
  {journal} {\bibinfo  {journal} {Rev. Mod. Phys.}\ }\textbf {\bibinfo {volume}
  {80}},\ \bibinfo {pages} {1083} (\bibinfo {year} {2008})}\BibitemShut
  {NoStop}%
\bibitem [{Wan()}]{Wang_book}%
  \BibitemOpen
  \href {\doibase 10.1090/cbms/112} {}\bibinfo {note} {{Z. Wang, {\it
  Topological Quantum Computation}, CBMS Regional Conference Series in
  Mathematics, No. 112 (American Mathematical Society, Providence, RI,
  2010)}}\BibitemShut {NoStop}%
\bibitem [{Pre()}]{Preskill_HP}%
  \BibitemOpen
  \href@noop {} {}\bibinfo {note} {See
  {\href{http://www.theory.caltech.edu/~preskill/ph219/}{http://www.theory.caltech.edu/$\sim$preskill/ph219/}
  for a pedagogical introduction}}\BibitemShut {NoStop}%
\bibitem [{\citenamefont {Witten}(1989)}]{Witten89}%
  \BibitemOpen
  \bibfield  {author} {\bibinfo {author} {\bibfnamefont {E.}~\bibnamefont
  {Witten}},\ }\bibfield  {title} {\enquote {\bibinfo {title} {{Quantum field
  theory and the Jones polynomial}},}\ }\href {\doibase 10.1007/BF01217730}
  {\bibfield  {journal} {\bibinfo  {journal} {Commun. Math. Phys.}\ }\textbf
  {\bibinfo {volume} {121}},\ \bibinfo {pages} {351} (\bibinfo {year}
  {1989})}\BibitemShut {NoStop}%
\bibitem [{\citenamefont {Levin}\ and\ \citenamefont {Wen}(2005)}]{Levin05}%
  \BibitemOpen
  \bibfield  {author} {\bibinfo {author} {\bibfnamefont {M.~A.}\ \bibnamefont
  {Levin}}\ and\ \bibinfo {author} {\bibfnamefont {X.-G.}\ \bibnamefont
  {Wen}},\ }\bibfield  {title} {\enquote {\bibinfo {title} {{String-net
  condensation: A physical mechanism for topological phases}},}\ }\href
  {\doibase 10.1103/PhysRevB.71.045110} {\bibfield  {journal} {\bibinfo
  {journal} {Phys. Rev. B}\ }\textbf {\bibinfo {volume} {71}},\ \bibinfo
  {pages} {045110} (\bibinfo {year} {2005})}\BibitemShut {NoStop}%
\bibitem [{\citenamefont {Hahn}\ and\ \citenamefont {Wolf}(2020)}]{Hahn20}%
  \BibitemOpen
  \bibfield  {author} {\bibinfo {author} {\bibfnamefont {A.}~\bibnamefont
  {Hahn}}\ and\ \bibinfo {author} {\bibfnamefont {R.}~\bibnamefont {Wolf}},\
  }\bibfield  {title} {\enquote {\bibinfo {title} {{Generalized string-net
  model for unitary fusion categories without tetrahedral symmetry}},}\ }\href
  {\doibase 10.1103/PhysRevB.102.115154} {\bibfield  {journal} {\bibinfo
  {journal} {Phys. Rev. B}\ }\textbf {\bibinfo {volume} {102}},\ \bibinfo
  {pages} {115154} (\bibinfo {year} {2020})}\BibitemShut {NoStop}%
\bibitem [{\citenamefont {Lin}\ \emph {et~al.}(2021)\citenamefont {Lin},
  \citenamefont {Levin},\ and\ \citenamefont {Burnell}}]{Lin21}%
  \BibitemOpen
  \bibfield  {author} {\bibinfo {author} {\bibfnamefont {C.-H.}\ \bibnamefont
  {Lin}}, \bibinfo {author} {\bibfnamefont {M.}~\bibnamefont {Levin}}, \ and\
  \bibinfo {author} {\bibfnamefont {F.~J.}\ \bibnamefont {Burnell}},\
  }\bibfield  {title} {\enquote {\bibinfo {title} {{Generalized string-net
  models: A thorough exposition}},}\ }\href {\doibase
  10.1103/PhysRevB.103.195155} {\bibfield  {journal} {\bibinfo  {journal}
  {Phys. Rev. B}\ }\textbf {\bibinfo {volume} {103}},\ \bibinfo {pages}
  {195155} (\bibinfo {year} {2021})}\BibitemShut {NoStop}%
\bibitem [{\citenamefont {Wolf}()}]{Wolf_thesis}%
  \BibitemOpen
  \bibfield  {author} {\bibinfo {author} {\bibfnamefont {R.}~\bibnamefont
  {Wolf}},\ }\href@noop {} {}\bibinfo {note}
  {\href{https://doi.org/10.15488/10324}{Ph. D. thesis, Leibniz Universit\"{a}t
  Hannover, 2020}}\BibitemShut {NoStop}%
\bibitem [{\citenamefont {Gils}\ \emph {et~al.}(2009)\citenamefont {Gils},
  \citenamefont {Trebst}, \citenamefont {Kitaev}, \citenamefont {Ludwig},
  \citenamefont {Troyer},\ and\ \citenamefont {Wang}}]{Gils09_1}%
  \BibitemOpen
  \bibfield  {author} {\bibinfo {author} {\bibfnamefont {C.}~\bibnamefont
  {Gils}}, \bibinfo {author} {\bibfnamefont {S.}~\bibnamefont {Trebst}},
  \bibinfo {author} {\bibfnamefont {A.}~\bibnamefont {Kitaev}}, \bibinfo
  {author} {\bibfnamefont {A.~W.~W.}\ \bibnamefont {Ludwig}}, \bibinfo {author}
  {\bibfnamefont {M.}~\bibnamefont {Troyer}}, \ and\ \bibinfo {author}
  {\bibfnamefont {Z.}~\bibnamefont {Wang}},\ }\bibfield  {title} {\enquote
  {\bibinfo {title} {{Topology-driven quantum phase transitions in
  time-reversal-invariant anyonic quantum liquids}},}\ }\href {\doibase
  10.1038/nphys1396} {\bibfield  {journal} {\bibinfo  {journal} {Nat. Phys.}\
  }\textbf {\bibinfo {volume} {5}},\ \bibinfo {pages} {834} (\bibinfo {year}
  {2009})}\BibitemShut {NoStop}%
\bibitem [{\citenamefont {Gils}()}]{Gils09_3}%
  \BibitemOpen
  \bibfield  {author} {\bibinfo {author} {\bibfnamefont {C.}~\bibnamefont
  {Gils}},\ }\href@noop {} {\enquote {\bibinfo {title} {{Ashkin-Teller
  universality in a quantum double model of Ising anyons}},}\ }\bibinfo {note}
  {\href{http://iopscience.iop.org/1742-5468/2009/07/P07019/}{J. Stat. Mech.,
  P07019 (2009)}}\BibitemShut {NoStop}%
\bibitem [{\citenamefont {Ardonne}\ \emph {et~al.}(2011)\citenamefont
  {Ardonne}, \citenamefont {Gukelberger}, \citenamefont {Ludwig}, \citenamefont
  {Trebst},\ and\ \citenamefont {Troyer}}]{Ardonne11}%
  \BibitemOpen
  \bibfield  {author} {\bibinfo {author} {\bibfnamefont {E.}~\bibnamefont
  {Ardonne}}, \bibinfo {author} {\bibfnamefont {J.}~\bibnamefont
  {Gukelberger}}, \bibinfo {author} {\bibfnamefont {A.~W.~W.}\ \bibnamefont
  {Ludwig}}, \bibinfo {author} {\bibfnamefont {S.}~\bibnamefont {Trebst}}, \
  and\ \bibinfo {author} {\bibfnamefont {M.}~\bibnamefont {Troyer}},\
  }\bibfield  {title} {\enquote {\bibinfo {title} {{Microscopic models of
  interacting Yang-Lee anyons}},}\ }\href {\doibase
  10.1088/1367-2630/13/4/045006} {\bibfield  {journal} {\bibinfo  {journal}
  {New J. Phys.}\ }\textbf {\bibinfo {volume} {13}},\ \bibinfo {pages} {045006}
  (\bibinfo {year} {2011})}\BibitemShut {NoStop}%
\bibitem [{\citenamefont {Burnell}\ \emph {et~al.}(2011)\citenamefont
  {Burnell}, \citenamefont {Simon},\ and\ \citenamefont
  {Slingerland}}]{Burnell11_2}%
  \BibitemOpen
  \bibfield  {author} {\bibinfo {author} {\bibfnamefont {F.~J.}\ \bibnamefont
  {Burnell}}, \bibinfo {author} {\bibfnamefont {S.~H.}\ \bibnamefont {Simon}},
  \ and\ \bibinfo {author} {\bibfnamefont {J.~K.}\ \bibnamefont
  {Slingerland}},\ }\bibfield  {title} {\enquote {\bibinfo {title}
  {{Condensation of achiral simple currents in topological lattice models:
  Hamiltonian study of topological symmetry breaking}},}\ }\href {\doibase
  10.1103/PhysRevB.84.125434} {\bibfield  {journal} {\bibinfo  {journal} {Phys.
  Rev. B}\ }\textbf {\bibinfo {volume} {84}},\ \bibinfo {pages} {125434}
  (\bibinfo {year} {2011})}\BibitemShut {NoStop}%
\bibitem [{\citenamefont {Schulz}\ \emph {et~al.}(2013)\citenamefont {Schulz},
  \citenamefont {Dusuel}, \citenamefont {Schmidt},\ and\ \citenamefont
  {Vidal}}]{Schulz13}%
  \BibitemOpen
  \bibfield  {author} {\bibinfo {author} {\bibfnamefont {M.~D.}\ \bibnamefont
  {Schulz}}, \bibinfo {author} {\bibfnamefont {S.}~\bibnamefont {Dusuel}},
  \bibinfo {author} {\bibfnamefont {K.~P.}\ \bibnamefont {Schmidt}}, \ and\
  \bibinfo {author} {\bibfnamefont {J.}~\bibnamefont {Vidal}},\ }\bibfield
  {title} {\enquote {\bibinfo {title} {{Topological Phase Transitions in the
  Golden String-Net Model}},}\ }\href {\doibase 10.1103/PhysRevLett.110.147203}
  {\bibfield  {journal} {\bibinfo  {journal} {Phys. Rev. Lett.}\ }\textbf
  {\bibinfo {volume} {110}},\ \bibinfo {pages} {147203} (\bibinfo {year}
  {2013})}\BibitemShut {NoStop}%
\bibitem [{\citenamefont {Schulz}\ \emph {et~al.}(2014)\citenamefont {Schulz},
  \citenamefont {Dusuel}, \citenamefont {Misguich}, \citenamefont {Schmidt},\
  and\ \citenamefont {Vidal}}]{Schulz14}%
  \BibitemOpen
  \bibfield  {author} {\bibinfo {author} {\bibfnamefont {M.~D.}\ \bibnamefont
  {Schulz}}, \bibinfo {author} {\bibfnamefont {S.}~\bibnamefont {Dusuel}},
  \bibinfo {author} {\bibfnamefont {G.}~\bibnamefont {Misguich}}, \bibinfo
  {author} {\bibfnamefont {K.~P.}\ \bibnamefont {Schmidt}}, \ and\ \bibinfo
  {author} {\bibfnamefont {J.}~\bibnamefont {Vidal}},\ }\bibfield  {title}
  {\enquote {\bibinfo {title} {{Ising anyons with a string tension}},}\ }\href
  {\doibase 10.1103/PhysRevB.89.201103} {\bibfield  {journal} {\bibinfo
  {journal} {Phys. Rev. B}\ }\textbf {\bibinfo {volume} {89}},\ \bibinfo
  {pages} {201103(R)} (\bibinfo {year} {2014})}\BibitemShut {NoStop}%
\bibitem [{\citenamefont {Dusuel}\ and\ \citenamefont
  {Vidal}(2015)}]{Dusuel15}%
  \BibitemOpen
  \bibfield  {author} {\bibinfo {author} {\bibfnamefont {S.}~\bibnamefont
  {Dusuel}}\ and\ \bibinfo {author} {\bibfnamefont {J.}~\bibnamefont {Vidal}},\
  }\bibfield  {title} {\enquote {\bibinfo {title} {{Mean-field ansatz for
  topological phases with string tension}},}\ }\href {\doibase
  10.1103/PhysRevB.92.125150} {\bibfield  {journal} {\bibinfo  {journal} {Phys.
  Rev. B}\ }\textbf {\bibinfo {volume} {92}},\ \bibinfo {pages} {125150}
  (\bibinfo {year} {2015})}\BibitemShut {NoStop}%
\bibitem [{\citenamefont {Schulz}\ \emph {et~al.}(2015)\citenamefont {Schulz},
  \citenamefont {Dusuel},\ and\ \citenamefont {Vidal}}]{Schulz15}%
  \BibitemOpen
  \bibfield  {author} {\bibinfo {author} {\bibfnamefont {M.~D.}\ \bibnamefont
  {Schulz}}, \bibinfo {author} {\bibfnamefont {S.}~\bibnamefont {Dusuel}}, \
  and\ \bibinfo {author} {\bibfnamefont {J.}~\bibnamefont {Vidal}},\ }\bibfield
   {title} {\enquote {\bibinfo {title} {{Russian doll spectrum in a non-Abelian
  string-net ladder}},}\ }\href {\doibase 10.1103/PhysRevB.91.155110}
  {\bibfield  {journal} {\bibinfo  {journal} {Phys. Rev. B}\ }\textbf {\bibinfo
  {volume} {91}},\ \bibinfo {pages} {155110} (\bibinfo {year}
  {2015})}\BibitemShut {NoStop}%
\bibitem [{\citenamefont {Schulz}\ \emph {et~al.}(2016)\citenamefont {Schulz},
  \citenamefont {Dusuel},\ and\ \citenamefont {Vidal}}]{Schulz16_2}%
  \BibitemOpen
  \bibfield  {author} {\bibinfo {author} {\bibfnamefont {M.~D.}\ \bibnamefont
  {Schulz}}, \bibinfo {author} {\bibfnamefont {S.}~\bibnamefont {Dusuel}}, \
  and\ \bibinfo {author} {\bibfnamefont {J.}~\bibnamefont {Vidal}},\ }\bibfield
   {title} {\enquote {\bibinfo {title} {{Bound states in string nets }},}\
  }\href {\doibase 10.1103/PhysRevB.94.205102} {\bibfield  {journal} {\bibinfo
  {journal} {Phys. Rev. B}\ }\textbf {\bibinfo {volume} {94}},\ \bibinfo
  {pages} {205102} (\bibinfo {year} {2016})}\BibitemShut {NoStop}%
\bibitem [{\citenamefont {Mari{\"{e}}n}\ \emph {et~al.}(2017)\citenamefont
  {Mari{\"{e}}n}, \citenamefont {Haegeman}, \citenamefont {Fendley},\ and\
  \citenamefont {Verstraete}}]{Mariens17}%
  \BibitemOpen
  \bibfield  {author} {\bibinfo {author} {\bibfnamefont {M.}~\bibnamefont
  {Mari{\"{e}}n}}, \bibinfo {author} {\bibfnamefont {J.}~\bibnamefont
  {Haegeman}}, \bibinfo {author} {\bibfnamefont {P.}~\bibnamefont {Fendley}}, \
  and\ \bibinfo {author} {\bibfnamefont {F.}~\bibnamefont {Verstraete}},\
  }\bibfield  {title} {\enquote {\bibinfo {title} {{Condensation-driven phase
  transitions in perturbed string nets}},}\ }\href {\doibase
  10.1103/PhysRevB.96.155127} {\bibfield  {journal} {\bibinfo  {journal} {Phys.
  Rev. B}\ }\textbf {\bibinfo {volume} {96}},\ \bibinfo {pages} {155127}
  (\bibinfo {year} {2017})}\BibitemShut {NoStop}%
\bibitem [{\citenamefont {Schotte}\ \emph {et~al.}(2019)\citenamefont
  {Schotte}, \citenamefont {Carrasco}, \citenamefont {Vanhecke}, \citenamefont
  {Vanderstraeten}, \citenamefont {Haegeman}, \citenamefont {Verstraete},\ and\
  \citenamefont {Vidal}}]{Schotte19}%
  \BibitemOpen
  \bibfield  {author} {\bibinfo {author} {\bibfnamefont {A.}~\bibnamefont
  {Schotte}}, \bibinfo {author} {\bibfnamefont {J.}~\bibnamefont {Carrasco}},
  \bibinfo {author} {\bibfnamefont {B.}~\bibnamefont {Vanhecke}}, \bibinfo
  {author} {\bibfnamefont {L.}~\bibnamefont {Vanderstraeten}}, \bibinfo
  {author} {\bibfnamefont {J.}~\bibnamefont {Haegeman}}, \bibinfo {author}
  {\bibfnamefont {F.}~\bibnamefont {Verstraete}}, \ and\ \bibinfo {author}
  {\bibfnamefont {J.}~\bibnamefont {Vidal}},\ }\bibfield  {title} {\enquote
  {\bibinfo {title} {{Tensor-network approach to phase transitions in
  string-net models}},}\ }\href {\doibase 10.1103/PhysRevB.100.245125}
  {\bibfield  {journal} {\bibinfo  {journal} {Phys. Rev. B}\ }\textbf {\bibinfo
  {volume} {100}},\ \bibinfo {pages} {245125} (\bibinfo {year}
  {2019})}\BibitemShut {NoStop}%
\bibitem [{\citenamefont {Ritz-Zwilling}\ \emph {et~al.}(2021)\citenamefont
  {Ritz-Zwilling}, \citenamefont {Fuchs},\ and\ \citenamefont
  {Vidal}}]{Ritz21}%
  \BibitemOpen
  \bibfield  {author} {\bibinfo {author} {\bibfnamefont {A.}~\bibnamefont
  {Ritz-Zwilling}}, \bibinfo {author} {\bibfnamefont {J.-N.}\ \bibnamefont
  {Fuchs}}, \ and\ \bibinfo {author} {\bibfnamefont {J.}~\bibnamefont
  {Vidal}},\ }\bibfield  {title} {\enquote {\bibinfo {title} {{Wegner-Wilson
  loops in string nets}},}\ }\href {\doibase 10.1103/PhysRevB.103.075128}
  {\bibfield  {journal} {\bibinfo  {journal} {Phys. Rev. B}\ }\textbf {\bibinfo
  {volume} {103}},\ \bibinfo {pages} {075128} (\bibinfo {year}
  {2021})}\BibitemShut {NoStop}%
\bibitem [{\citenamefont {K\'ad\'ar}\ \emph {et~al.}(2010)\citenamefont
  {K\'ad\'ar}, \citenamefont {Marzuoli},\ and\ \citenamefont
  {M.Rasetti}}]{Kadar10}%
  \BibitemOpen
  \bibfield  {author} {\bibinfo {author} {\bibfnamefont {Z.}~\bibnamefont
  {K\'ad\'ar}}, \bibinfo {author} {\bibfnamefont {A.}~\bibnamefont {Marzuoli}},
  \ and\ \bibinfo {author} {\bibnamefont {M.Rasetti}},\ }\bibfield  {title}
  {\enquote {\bibinfo {title} {Microscopic description of 2d topological
  phases, duality, and 3d state sums},}\ }\href {\doibase 10.1155/2010/671039}
  {\bibfield  {journal} {\bibinfo  {journal} {Adv. Math. Phys.}\ }\textbf
  {\bibinfo {volume} {2010}},\ \bibinfo {pages} {671039} (\bibinfo {year}
  {2010})}\BibitemShut {NoStop}%
\bibitem [{\citenamefont {Burnell}\ and\ \citenamefont
  {Simon}(2010)}]{Burnell10}%
  \BibitemOpen
  \bibfield  {author} {\bibinfo {author} {\bibfnamefont {F.~J.}\ \bibnamefont
  {Burnell}}\ and\ \bibinfo {author} {\bibfnamefont {S.~H.}\ \bibnamefont
  {Simon}},\ }\bibfield  {title} {\enquote {\bibinfo {title} {Space-time
  geometry of topological phases},}\ }\href {\doibase
  10.1016/j.aop.2010.06.003} {\bibfield  {journal} {\bibinfo  {journal} {Ann.
  Phys. (NY)}\ }\textbf {\bibinfo {volume} {325}},\ \bibinfo {pages} {2550}
  (\bibinfo {year} {2010})}\BibitemShut {NoStop}%
\bibitem [{\citenamefont {Burnell}\ and\ \citenamefont
  {Simon}(2011)}]{Burnell11_1}%
  \BibitemOpen
  \bibfield  {author} {\bibinfo {author} {\bibfnamefont {F.~J.}\ \bibnamefont
  {Burnell}}\ and\ \bibinfo {author} {\bibfnamefont {S.~H.}\ \bibnamefont
  {Simon}},\ }\bibfield  {title} {\enquote {\bibinfo {title} {{A Wilson line
  picture of the Levin-Wen partition functions}},}\ }\href {\doibase
  10.1088/1367-2630/13/6/065001} {\bibfield  {journal} {\bibinfo  {journal}
  {New J. Phys.}\ }\textbf {\bibinfo {volume} {13}},\ \bibinfo {pages} {065001}
  (\bibinfo {year} {2011})}\BibitemShut {NoStop}%
\bibitem [{\citenamefont {Kirillov}()}]{Kirillov11}%
  \BibitemOpen
  \bibfield  {author} {\bibinfo {author} {\bibfnamefont {A.}~\bibnamefont
  {Kirillov}},\ }\href@noop {} {\enquote {\bibinfo {title} {{String-net model
  of Turaev-Viro invariants}},}\ }\bibinfo {note}
  {\href{https://arxiv.org/abs/1106.6033}{arXiv:1106.6033}}\BibitemShut
  {NoStop}%
\bibitem [{\citenamefont {Turaev}\ and\ \citenamefont {Viro}(1992)}]{Turaev92}%
  \BibitemOpen
  \bibfield  {author} {\bibinfo {author} {\bibfnamefont {V.~G.}\ \bibnamefont
  {Turaev}}\ and\ \bibinfo {author} {\bibfnamefont {O.~Y.}\ \bibnamefont
  {Viro}},\ }\bibfield  {title} {\enquote {\bibinfo {title} {{State sum
  invariants of 3-manifolds and quantum $6j$-symbols}},}\ }\href {\doibase
  10.1016/0040-9383(92)90015-A} {\bibfield  {journal} {\bibinfo  {journal}
  {Topology}\ }\textbf {\bibinfo {volume} {31}},\ \bibinfo {pages} {865}
  (\bibinfo {year} {1992})}\BibitemShut {NoStop}%
\bibitem [{\citenamefont {Reshetikhin}\ and\ \citenamefont
  {Turaev}(1991)}]{Reshetikhin91}%
  \BibitemOpen
  \bibfield  {author} {\bibinfo {author} {\bibfnamefont {N.}~\bibnamefont
  {Reshetikhin}}\ and\ \bibinfo {author} {\bibfnamefont {V.~G.}\ \bibnamefont
  {Turaev}},\ }\bibfield  {title} {\enquote {\bibinfo {title} {{Invariants of
  3-manifolds via link polynomials and quantum groups}},}\ }\href {\doibase
  10.1007/BF01239527} {\bibfield  {journal} {\bibinfo  {journal} {Invent.
  Math.}\ }\textbf {\bibinfo {volume} {103}},\ \bibinfo {pages} {547} (\bibinfo
  {year} {1991})}\BibitemShut {NoStop}%
\bibitem [{\citenamefont {Simon}\ and\ \citenamefont
  {Fendley}(2013)}]{Simon13}%
  \BibitemOpen
  \bibfield  {author} {\bibinfo {author} {\bibfnamefont {S.~H.}\ \bibnamefont
  {Simon}}\ and\ \bibinfo {author} {\bibfnamefont {P.}~\bibnamefont
  {Fendley}},\ }\bibfield  {title} {\enquote {\bibinfo {title} {{Exactly
  solvable lattice models with crossing symmetry}},}\ }\href {\doibase
  10.1088/1751-8113/46/10/105002} {\bibfield  {journal} {\bibinfo  {journal}
  {J. Phys. A: Math. Theor.}\ }\textbf {\bibinfo {volume} {46}},\ \bibinfo
  {pages} {105002} (\bibinfo {year} {2013})}\BibitemShut {NoStop}%
\bibitem [{\citenamefont {Hu}\ \emph {et~al.}(2018)\citenamefont {Hu},
  \citenamefont {Geer},\ and\ \citenamefont {Wu}}]{Hu18}%
  \BibitemOpen
  \bibfield  {author} {\bibinfo {author} {\bibfnamefont {Y.}~\bibnamefont
  {Hu}}, \bibinfo {author} {\bibfnamefont {N.}~\bibnamefont {Geer}}, \ and\
  \bibinfo {author} {\bibfnamefont {Y.-S.}\ \bibnamefont {Wu}},\ }\bibfield
  {title} {\enquote {\bibinfo {title} {{Full dyon excitation spectrum in
  extended Levin-Wen models}},}\ }\href {\doibase 10.1103/PhysRevB.97.195154}
  {\bibfield  {journal} {\bibinfo  {journal} {Phys. Rev. B}\ }\textbf {\bibinfo
  {volume} {97}},\ \bibinfo {pages} {195154} (\bibinfo {year}
  {2018})}\BibitemShut {NoStop}%
\bibitem [{\citenamefont {{S. C. Morampudi and C. von Keyserlingk and F.
  Pollmann}}(2014)}]{Morampudi14}%
  \BibitemOpen
  \bibfield  {author} {\bibinfo {author} {\bibnamefont {{S. C. Morampudi and C.
  von Keyserlingk and F. Pollmann}}},\ }\bibfield  {title} {\enquote {\bibinfo
  {title} {{Numerical study of a transition between $\mathbb{Z}_2$
  topologically-ordered phases}},}\ }\href {\doibase
  10.1103/PhysRevB.90.035117} {\bibfield  {journal} {\bibinfo  {journal} {Phys.
  Rev. B}\ }\textbf {\bibinfo {volume} {90}},\ \bibinfo {pages} {035117}
  (\bibinfo {year} {2014})}\BibitemShut {NoStop}%
\bibitem [{\citenamefont {Vidal}(2018)}]{Vidal18}%
  \BibitemOpen
  \bibfield  {author} {\bibinfo {author} {\bibfnamefont {J.}~\bibnamefont
  {Vidal}},\ }\bibfield  {title} {\enquote {\bibinfo {title} {{Ising versus
  $SU(2)_2$ string-net ladder}},}\ }\href {\doibase 10.1103/PhysRevB.125152}
  {\bibfield  {journal} {\bibinfo  {journal} {Phys. Rev. B}\ }\textbf {\bibinfo
  {volume} {97}},\ \bibinfo {pages} {125152} (\bibinfo {year}
  {2018})}\BibitemShut {NoStop}%
\bibitem [{\citenamefont {Rowell}\ \emph {et~al.}(2009)\citenamefont {Rowell},
  \citenamefont {Stong},\ and\ \citenamefont {Wang}}]{Rowell09}%
  \BibitemOpen
  \bibfield  {author} {\bibinfo {author} {\bibfnamefont {E.}~\bibnamefont
  {Rowell}}, \bibinfo {author} {\bibfnamefont {R.}~\bibnamefont {Stong}}, \
  and\ \bibinfo {author} {\bibfnamefont {Z.}~\bibnamefont {Wang}},\ }\bibfield
  {title} {\enquote {\bibinfo {title} {On classification of modular tensor
  categories},}\ }\href {\doibase 10.1007/s00220-009-0908-z} {\bibfield
  {journal} {\bibinfo  {journal} {Commun. Math. Phys.}\ }\textbf {\bibinfo
  {volume} {292}},\ \bibinfo {pages} {343} (\bibinfo {year}
  {2009})}\BibitemShut {NoStop}%
\bibitem [{\citenamefont {Bonderson}()}]{Bonderson_thesis}%
  \BibitemOpen
  \bibfield  {author} {\bibinfo {author} {\bibfnamefont {P.~H.}\ \bibnamefont
  {Bonderson}},\ }\href@noop {} {}\bibinfo {note}
  {\href{https://doi.org/10.7907/5NDZ-W890}{Ph. D. thesis, California Institute
  of Technology, 2007}}\BibitemShut {NoStop}%
\bibitem [{\citenamefont {Pasquier}(1987)}]{Pasquier87}%
  \BibitemOpen
  \bibfield  {author} {\bibinfo {author} {\bibfnamefont {V.}~\bibnamefont
  {Pasquier}},\ }\bibfield  {title} {\enquote {\bibinfo {title} {{Operator
  content of the ADE lattice models}},}\ }\href {\doibase
  10.1088/0305-4470/20/16/043} {\bibfield  {journal} {\bibinfo  {journal} {J.
  Phys. A}\ }\textbf {\bibinfo {volume} {20}},\ \bibinfo {pages} {5707}
  (\bibinfo {year} {1987})}\BibitemShut {NoStop}%
\bibitem [{\citenamefont {Verlinde}(1988)}]{Verlinde88}%
  \BibitemOpen
  \bibfield  {author} {\bibinfo {author} {\bibfnamefont {E.}~\bibnamefont
  {Verlinde}},\ }\bibfield  {title} {\enquote {\bibinfo {title} {{Fusion rules
  and modular transformations in 2D conformal field theory}},}\ }\href
  {\doibase 10.1016/0550-3213(88)90603-7} {\bibfield  {journal} {\bibinfo
  {journal} {Nucl. Phys. B}\ }\textbf {\bibinfo {volume} {300}},\ \bibinfo
  {pages} {360} (\bibinfo {year} {1988})}\BibitemShut {NoStop}%
\bibitem [{\citenamefont {Gannon}(2005)}]{Gannon05}%
  \BibitemOpen
  \bibfield  {author} {\bibinfo {author} {\bibfnamefont {T.}~\bibnamefont
  {Gannon}},\ }\bibfield  {title} {\enquote {\bibinfo {title} {Modular data:
  The algebraic combinatorics of conformal field theory},}\ }\href {\doibase
  10.1007/s10801-005-2514-2} {\bibfield  {journal} {\bibinfo  {journal} {J.
  Algebr. Comb.}\ }\textbf {\bibinfo {volume} {22}},\ \bibinfo {pages} {211}
  (\bibinfo {year} {2005})}\BibitemShut {NoStop}%
\bibitem [{\citenamefont {Barkeshli}\ and\ \citenamefont
  {Wen}(2009)}]{Barkeshli09}%
  \BibitemOpen
  \bibfield  {author} {\bibinfo {author} {\bibfnamefont {M.}~\bibnamefont
  {Barkeshli}}\ and\ \bibinfo {author} {\bibfnamefont {X.-G.}\ \bibnamefont
  {Wen}},\ }\bibfield  {title} {\enquote {\bibinfo {title} {{Structure of
  quasiparticles and their fusion algebra in fractional quantum Hall
  states}},}\ }\href {\doibase 10.1103/PhysRevB.79.195132} {\bibfield
  {journal} {\bibinfo  {journal} {Phys. Rev. B}\ }\textbf {\bibinfo {volume}
  {79}},\ \bibinfo {pages} {195132} (\bibinfo {year} {2009})}\BibitemShut
  {NoStop}%
\bibitem [{\citenamefont {Hu}\ \emph {et~al.}(2012)\citenamefont {Hu},
  \citenamefont {Stirling},\ and\ \citenamefont {Wu}}]{Hu12}%
  \BibitemOpen
  \bibfield  {author} {\bibinfo {author} {\bibfnamefont {Y.}~\bibnamefont
  {Hu}}, \bibinfo {author} {\bibfnamefont {S.~D.}\ \bibnamefont {Stirling}}, \
  and\ \bibinfo {author} {\bibfnamefont {Y.-S.}\ \bibnamefont {Wu}},\
  }\bibfield  {title} {\enquote {\bibinfo {title} {{Ground-state degeneracy in
  the Levin-Wen model for topological phases}},}\ }\href {\doibase
  10.1103/PhysRevB.85.075107} {\bibfield  {journal} {\bibinfo  {journal} {Phys.
  Rev. B}\ }\textbf {\bibinfo {volume} {85}},\ \bibinfo {pages} {075107}
  (\bibinfo {year} {2012})}\BibitemShut {NoStop}%
\bibitem [{Dri()}]{Drinfeld87}%
  \BibitemOpen
  \href {\doibase 10.1090/cbms/112} {}\bibinfo {note} {{V. G. Drinfel'd, {\it
  Quantum groups}, Proceedings of the International Congress of Mathematicians,
  Vol. 1, 2 (Berkeley, CA, 1986), 798 (American Mathematical Society,
  Providence, RI, 1987)}}\BibitemShut {NoStop}%
\bibitem [{Sup()}]{Supp_Mat}%
  \BibitemOpen
  \href@noop {} {}\bibinfo {note} {{See Supplemental Material.}}\BibitemShut
  {Stop}%
\bibitem [{\citenamefont {Hu}\ \emph {et~al.}(2014)\citenamefont {Hu},
  \citenamefont {Stirling},\ and\ \citenamefont {Wu}}]{Hu14}%
  \BibitemOpen
  \bibfield  {author} {\bibinfo {author} {\bibfnamefont {Y.}~\bibnamefont
  {Hu}}, \bibinfo {author} {\bibfnamefont {S.~D.}\ \bibnamefont {Stirling}}, \
  and\ \bibinfo {author} {\bibfnamefont {Y.-S.}\ \bibnamefont {Wu}},\
  }\bibfield  {title} {\enquote {\bibinfo {title} {{Emergent exclusion
  statistics of quasiparticles in two-dimensional topological phases}},}\
  }\href {\doibase 10.1103/PhysRevB.89.115133} {\bibfield  {journal} {\bibinfo
  {journal} {Phys. Rev. B}\ }\textbf {\bibinfo {volume} {89}},\ \bibinfo
  {pages} {115133} (\bibinfo {year} {2014})}\BibitemShut {NoStop}%
\bibitem [{Gil()}]{Gils09_1_s}%
  \BibitemOpen
  \href@noop {} {}\bibinfo {note} {{Supplementary information from
  \cite{Gils09_1}.}}\BibitemShut {Stop}%
\bibitem [{\citenamefont {Hermanns}\ and\ \citenamefont
  {Trebst}(2014)}]{Hermanns14}%
  \BibitemOpen
  \bibfield  {author} {\bibinfo {author} {\bibfnamefont {M.}~\bibnamefont
  {Hermanns}}\ and\ \bibinfo {author} {\bibfnamefont {S.}~\bibnamefont
  {Trebst}},\ }\bibfield  {title} {\enquote {\bibinfo {title} {{Renyi entropies
  for classical string-net models}},}\ }\href {\doibase
  10.1103/PhysRevB.89.205107} {\bibfield  {journal} {\bibinfo  {journal} {Phys.
  Rev. B}\ }\textbf {\bibinfo {volume} {89}},\ \bibinfo {pages} {205107}
  (\bibinfo {year} {2014})}\BibitemShut {NoStop}%
\bibitem [{\citenamefont {Weinstein}\ \emph {et~al.}(2019)\citenamefont
  {Weinstein}, \citenamefont {Ortiz},\ and\ \citenamefont
  {Nussinov}}]{Weinstein19}%
  \BibitemOpen
  \bibfield  {author} {\bibinfo {author} {\bibfnamefont {Z.}~\bibnamefont
  {Weinstein}}, \bibinfo {author} {\bibfnamefont {G.}~\bibnamefont {Ortiz}}, \
  and\ \bibinfo {author} {\bibfnamefont {Z.}~\bibnamefont {Nussinov}},\
  }\bibfield  {title} {\enquote {\bibinfo {title} {{Universality Classes of
  Stabilizer Code Hamiltonians}},}\ }\href {\doibase
  10.1103/PhysRevLett.123.230503} {\bibfield  {journal} {\bibinfo  {journal}
  {Phys. Rev. Lett.}\ }\textbf {\bibinfo {volume} {123}},\ \bibinfo {pages}
  {230503} (\bibinfo {year} {2019})}\BibitemShut {NoStop}%
\end{thebibliography}

%

\newpage

\onecolumngrid

\centerline{\bf \large SUPPLEMENTAL MATERIAL}

\appendix

\section{Derivation of Eq.~(\ref{eq:deg_g})}

In this Appendix, we give the key steps needed to derive Eq.~(\ref{eq:deg_g}) from Eq.~(\ref{eq:deg_f}) given in the main text. The degeneracy of the $q$-fluxon energy level is given by Eq.~(\ref{eq:deg_f}):
%
%
\begin{equation}
\mathcal{D}_q=
\left(\hspace{-1mm}
\begin{array}{c}
N_{\mathrm p}
\\
q
\end{array}
\hspace{-1mm}
\right) \sum_{\{ l_i\}/q}
\sum_{k_1=1}^n \sum_{k_2=1}^n 
\mathcal{F}^{\{l_i\}}_{k_1} \mathcal{F}^{\{l_i\}}_{k_2} \mathcal{G}_{g,k_1} \mathcal{G}_{g,k_2}
\left(\hspace{-1mm}\begin{array}{c}q \\l_2\end{array}\hspace{-1mm}\right) 
\left(\hspace{-1mm}\begin{array}{c}q-l_2 \\l_3\end{array}\hspace{-1mm}\right)\dots
 \left(\hspace{-1mm}\begin{array}{c}q-l_2-l_3-... \, l_{n-1} \\ l_n\end{array}\hspace{-1mm}\right). 
 \nonumber
\end{equation} 
%
%

In this expression, the first binomial takes into account all the possibilities to choose the position of the $q$ nontrivial fluxons among the $N_{\rm p}$ plaquettes. 
The product of binomials accounts for all possible permutations of the different strings in the set of these $q$ nontrivial fluxons. The first sum is performed over all possible choices of $\{ l_i\}$ such that $\sum_{i=2}^n l_i=q$ (here the case $i=1$ is discarded ($l_1=0$) since it does not correspond to an excitation).  The last two sums are performed over the possible values of $k_1$ and $k_2$ which are the inner and outer fluxons, respectively. Replacing $\mathcal{F}^{\{l_i\}}_{k}$  by its expression [see Eq.~(\ref{eq:fusion}) in the main text]  one gets:
%
%
\begin{equation}
\mathcal{D}_q=
\left(\hspace{-1mm}
\begin{array}{c}
N_{\mathrm p}
\\
q
\end{array}
\hspace{-1mm}
\right) \sum_{\{ l_i\}/q}
\sum_{k_1=1}^n \sum_{k_2=1}^n 
 \sum_{j_1=1}^{n}  \sum_{j_2=1}^{n}
 \left[S_{1,j_1} \prod_{i=2}^{n} \left( \frac{S_{i,j_1}}{S_{1,j_1}} \right)^{l_i} S^\dagger_{j_1,k_1} \right]
 \left[S_{1,j_2} \prod_{i=2}^{n} \left( \frac{S_{i,j_2}}{S_{1,j_2}} \right)^{l_i} S^\dagger_{j_2,k_2} \right]
\mathcal{G}_{g,k_1} \mathcal{G}_{g,k_2} \frac{q!}{\prod_{i=2}^{n} l_i!}
. 
\end{equation} 
%
%

As a first step, we perform the sum over $\{ l_i\}$ such that $\sum_{i=2}^n l_i=q$ since:
%
%
\begin{equation}
\sum_{\{ l_i\}/q}
 \prod_{i=2}^{n} \left( \frac{S_{i,j_1}}{S_{1,j_1}} \right)^{l_i} \prod_{i=2}^{n} \left( \frac{S_{i,j_2}}{S_{1,j_2}} \right)^{l_i}
\frac{q!}{\prod_{i=2}^{n} l_i!}=\sum_{\{ l_i\}/q}
 \prod_{i=2}^{n} \left( \frac{S_{i,j_1}}{S_{1,j_1}} \frac{S_{i,j_2}}{S_{1,j_2}}\right)^{l_i}
\frac{q!}{\prod_{i=2}^{n} l_i!}
= \left(\sum_{i=2}^n \frac{S_{i,j_1}}{S_{1,j_1}} \frac{S_{i,j_2}}{S_{1,j_2}}\right)^q.
\end{equation} 
%
%
Then, replacing $\mathcal{G}_{g,k}$ by its expression [see Eq.~(\ref{eq:G}) in the main text], one obtains:
%
%
\begin{equation}
\mathcal{D}_q=
\left(\hspace{-1mm}
\begin{array}{c}
N_{\mathrm p}
\\
q
\end{array}
\hspace{-1mm}
\right) 
\sum_{k_1=1}^n \sum_{k_2=1}^n 
\sum_{j_1=1}^n  \sum_{j_2=1}^n
\sum_{i_1=1}^n \sum_{i_2=1}^n 
S_{1,j_1}  S^\dagger_{j_1,k_1} 
S_{1,j_2}  S^\dagger_{j_2,k_2}  \left(\sum_{i=2}^n \frac{S_{i,j_1}}{S_{1,j_1}} \frac{S_{i,j_2}}{S_{1,j_2}}\right)^q
S_{1,i_1}^{1-2g} S^\dagger_{i_1,k_1} S_{1,i_2}^{1-2g} S^\dagger_{i_2,k_2}
. 
\end{equation} 
%
%
The second step consists in using the fact that $S$ is symmetric and that $(S^2)_{j,k}=1$ if $j$ and $k$ can fuse into the vacuum (i.e., they are conjugate strings), and $0$ otherwise \cite{Rowell09}. Since properties obviously hold for $S^\dagger$,  one has $\sum_{k=1}^n S^\dagger_{i,k} S^\dagger_{j,k}=\delta_{i,j}$, so that performing the summation over $k_1$ and $k_2$, one gets:
%
%
\begin{eqnarray}
\mathcal{D}_q&=&
\left(\hspace{-1mm}
\begin{array}{c}
N_{\mathrm p}
\\
q
\end{array}
\hspace{-1mm}
\right) 
\sum_{j_1=1}^n  \sum_{j_2=1}^n
\sum_{i_1=1}^n \sum_{i_2=1}^n 
S_{1,j_1}  
S_{1,j_2}   \left(\sum_{i=2}^n \frac{S_{i,j_1}}{S_{1,j_1}} \frac{S_{i,j_2}}{S_{1,j_2}}\right)^q
S_{1,i_1}^{1-2g} S_{1,i_2}^{1-2g}\delta_{i_1,j_1}  \delta_{i_2,j_2},\\
&=&
\left(\hspace{-1mm}
\begin{array}{c}
N_{\mathrm p}
\\
q
\end{array}
\hspace{-1mm}
\right) 
\sum_{j_1=1}^n  \sum_{j_2=1}^n S_{1,j_1}^{2-2g-q} S_{1,j_2}^{2-2g-q}
\left(\sum_{i=2}^n S_{i,j_1} S_{i,j_2} \right)^q.
\end{eqnarray} 
%
%
 Along the same line, one can perform the summation over $i$ by taking care of the fact that $i$ runs from $2$ to $n$ (instead of $1$ to $n$ for $k_1$ and $k_2$). This leads to:
 %
%
\begin{eqnarray}
\mathcal{D}_q&=&
\left(\hspace{-1mm}
\begin{array}{c}
N_{\mathrm p}
\\
q
\end{array}
\hspace{-1mm}
\right) 
\sum_{j_1=1}^n  \sum_{j_2=1}^n S_{1,j_1}^{\chi-q} S_{1,j_2}^{\chi-q}
\left(\delta_{j_1,j_2}- S_{1,j_1} S_{1,j_2} \right)^q,\\
&=& \left(\hspace{-1mm}
\begin{array}{c}
N_{\mathrm p}
\\
q
\end{array}
\hspace{-1mm}
\right) \left\{
(-1)^q \left[\left(\sum_{j=1}^n   S_{1,j}^{\chi} \right)^2 - \sum_{j=1}^n   S_{1,j}^{2 \chi}\right]+
\sum_{j=1}^n S_{1,j}^{2(\chi-q)} \left(1- S_{1,j}^2 \right)^q\right\},
\end{eqnarray} 
%
%
where we introduced the Euler-Poincar\'e characteristic $\chi=2-2g$. Finally, noting that, for a compact surface,  $\mathcal{D}_0=\left(\sum_{j=1}^n   S_{1,j}^{\chi} \right)^2$ [see Eq.~(\ref{eq:gse_deg})  in the main text], one gets the simple form given in Eq.~(\ref{eq:deg_g}) of the main text:
%
%
\begin{equation}
\mathcal{D}_q=
\left(\hspace{-1mm}
\begin{array}{c}
N_{\mathrm p}
\\
q
\end{array}
\hspace{-1mm}
\right) (-1)^q  \Bigg\{ {\mathcal D}_0+\sum_{j=1}^n S_{1,j}^{2 \chi}  \bigg[\Big(1-S_{1,j}^{-2}\Big)^q-1 \bigg]\Bigg\}.
\nonumber
\end{equation} 
%
%

\section{Hilbert space and degeneracies for the Fibonacci theory}

In this Appendix, we give a few details about the construction of the Hilbert space of the Levin-Wen model for the simplest non-Abelian  unitary modular tensor category (UMTC), i.e., the Fibonacci theory. Then, degeneracies of the energy levels are given for a few simple two-dimensional surfaces. 

\subsection{Fusion and $S$-matrices}

The Fibonacci UMTC contains two different labels, $1$ and $\tau$, which obey the following fusion rules (see e.g., Ref.~\onlinecite{Rowell09} for more details):
%
%
\begin{equation}
1\times1=1, \: 
1 \times \tau = \tau \times 1= \tau, \: 
\tau \times \tau= 1+\tau.
\end{equation} 
%
%
 To build the full Hilbert space of the Levin-Wen model on a given trivalent graph $G$, we assign to each link the state $1$ or $\tau$ so that one has, a priori, the Hilbert space dimension $2^{N_\mathrm{l}}$, where $N_\mathrm{l}$ is the total number of links of $G$. However, in the present study, we work in a restricted  Hilbert space in which the vertex constraint is satisfied (no charge excitation). This vertex constraint is imposed by requiring that each state satisfies the branching rules at each vertex of $G$. These branching rules are directly obtained from the fusion rules. More precisely, a vertex configuration $(a,b,c)$ is allowed iff the  $c$ belongs to the fusion product of $a \times b$. Hence, for the Fibonacci theory, all triplets are $(a,b,c)$ are allowed except $(1,1,\tau)$, $(1,\tau,1)$, and $(\tau,1,1)$.
From the fusion rules, one can directly obtain the quantum dimensions of the labels (real positive numbers obeying the fusion rules) $d_1=1$ and $d_\tau=\varphi$, where $\varphi=\frac{1+\sqrt{5}}{2}$ is the golden ratio. One can also build the two fusion matrices :
%
%
\begin{equation}
N_1=\left(\hspace{-1mm}
\begin{array}{c c} 
1 & 0 \\
0 & 1
\end{array}
\hspace{-1mm}\right), 
N_\tau=\left(\hspace{-1mm}\begin{array}{c c} 
0& 1\\
1& 1
\end{array}
\hspace{-1mm}\right).
%
%
\end{equation}

The $S$-matrix is a symmetric unitary  matrix which diagonalizes $N_1$ and $N_\tau$ simultaneously. Imposing the normalization constraint $S_{1,j}=d_j/D$, one gets \cite{Rowell09}:
%
%
\begin{equation}
S=\frac{1}{\sqrt{1+\varphi^2}}
\left(\hspace{-1mm}
\begin{array}{c c} 
1 & \varphi \\
\varphi & -1
\end{array}
\hspace{-1mm}\right),
%
%
\end{equation}
where $D=\sqrt{1+\varphi^2}$ is the total quantum dimension. This is all we need to compute the spectrum degeneracies.

\subsection{Degeneracies of the energy levels}

Equipped with the $S$-matrix, we can use Eq.~(\ref{eq:gse_deg})  and Eq.~(\ref{eq:deg_g}) given the main text to compute the degeneracies of the energy levels for various topologies. In Table I, we give the reduced degeneracy $\mathcal{D}_q\bigg/ \left(\hspace{-1mm}\begin{array}{c} N_\mathrm{p} \\q \end{array}\hspace{-1mm}\right)$ 
of the $q$-fluxon energy level ($E_q=-N_\mathrm{p}+q$, where $N_\mathrm{p}$ is the total number of plaquettes of $G$). This reduced degeneracy only depends on the Euler-Poincar\'e characteristic \mbox{$\chi=N_{\rm v}-N_{\rm l}+N_{\rm p}$}, where $N_{\rm v}$ $(N_{\rm l})$ denotes the total number of vertices (links) of $G$. For any trivalent graph, one has $N_{\rm l}=3 N_{\rm v}/2$. 
%
%
\begin{table}[h!]
\begin{tabular}{|c|c|c|c|c|c|}
\hline
$E_q-E_0$ & sphere ($\chi=2$) & 1-holed torus ($\chi=0$)& 2-holed torus ($\chi=-2$) & cylinder ($\chi=0$)& disk ($\chi=1$)\\
\hline
0  & 1&4&25& 2 & 1\\
1  & 0&1&25& 3 &1\\
2  & 1&9&100 &7 & 2\\
3  & 1&16&225&18 &5\\
4  & 4&49&625&47 &13\\
\hline
\end{tabular}
\caption{Reduced degeneracies of the first energy levels ($q=0,\dots, 4$) for different topologies. For non-compact surfaces (the cylinder and the disk), Eq.~(\ref{eq:gse_deg}) cannot be used and one must compute $\mathcal{D}_0$ separately using the same line of reasoning as the one presented in the main text. However, Eq.~(\ref{eq:deg_g}) remains valid.}
\label{tablemass}
\end{table}
%
%

\end{document}